 \documentclass[aps,prl,notitlepage,superscriptaddress,twocolumn ]{revtex4}
\usepackage{graphicx,subfigure,epsfig}
\usepackage{dcolumn}
\usepackage{amssymb}
\usepackage{times}
\usepackage{amsmath}
\usepackage{amsfonts}
\usepackage{mathrsfs}
\usepackage{setspace}
\usepackage{latexsym}
\usepackage{bbm}
\usepackage{float}
\usepackage{flafter}
\usepackage{bm}
\usepackage{epstopdf}
\usepackage{hyperref}
\usepackage{color}
\usepackage{multirow}

\def \be {\begin{eqnarray}}
\def \ee {\end{eqnarray}}

\begin{document}

\title{Widely Tunable Quantum Phase Transition from Moore-Read to Composite Fermi Liquid in Bilayer Graphene}

\author{Zheng Zhu}
\affiliation{Department of Physics, Harvard University, Cambridge, MA, 02138, USA}
\author{D. N. Sheng}
\affiliation{Department of Physics and Astronomy, California State University, Northridge, CA, 91330, USA}
\author{Inti Sodemann}
\affiliation{Max-Planck Institute for the Physics of Complex Systems, D-01187 Dresden, Germany}

\begin{abstract}
We develop a proposal to realise a widely tunable and clean quantum phase transition in bilayer graphene between two paradigmatic fractionalized phases of matter: the Moore-Read fractional quantum Hall state and the composite Fermi liquid metal. This transition can be realized at total fillings $\nu=\pm 3+1/2$ and the critical point can be controllably accessed by tuning either the interlayer electric bias or the perpendicular magnetic field values over a wide range of parameters. We study the transition numerically within a model that contains all leading single particle corrections to the band-structure of bilayer graphene and includes the fluctuations between the $n=0$ and $n=1$ cyclotron orbitals of its zeroth Landau level to delineate the most favorable region of parameters to experimentally access this unconventional critical point. We also find evidence for a new anisotropic gapless phase stabilized near the level crossing of $n=0/1$ orbits.
\end{abstract}

\maketitle

\begin{figure}[!t]
\begin{center}
\includegraphics[width=0.5\textwidth]{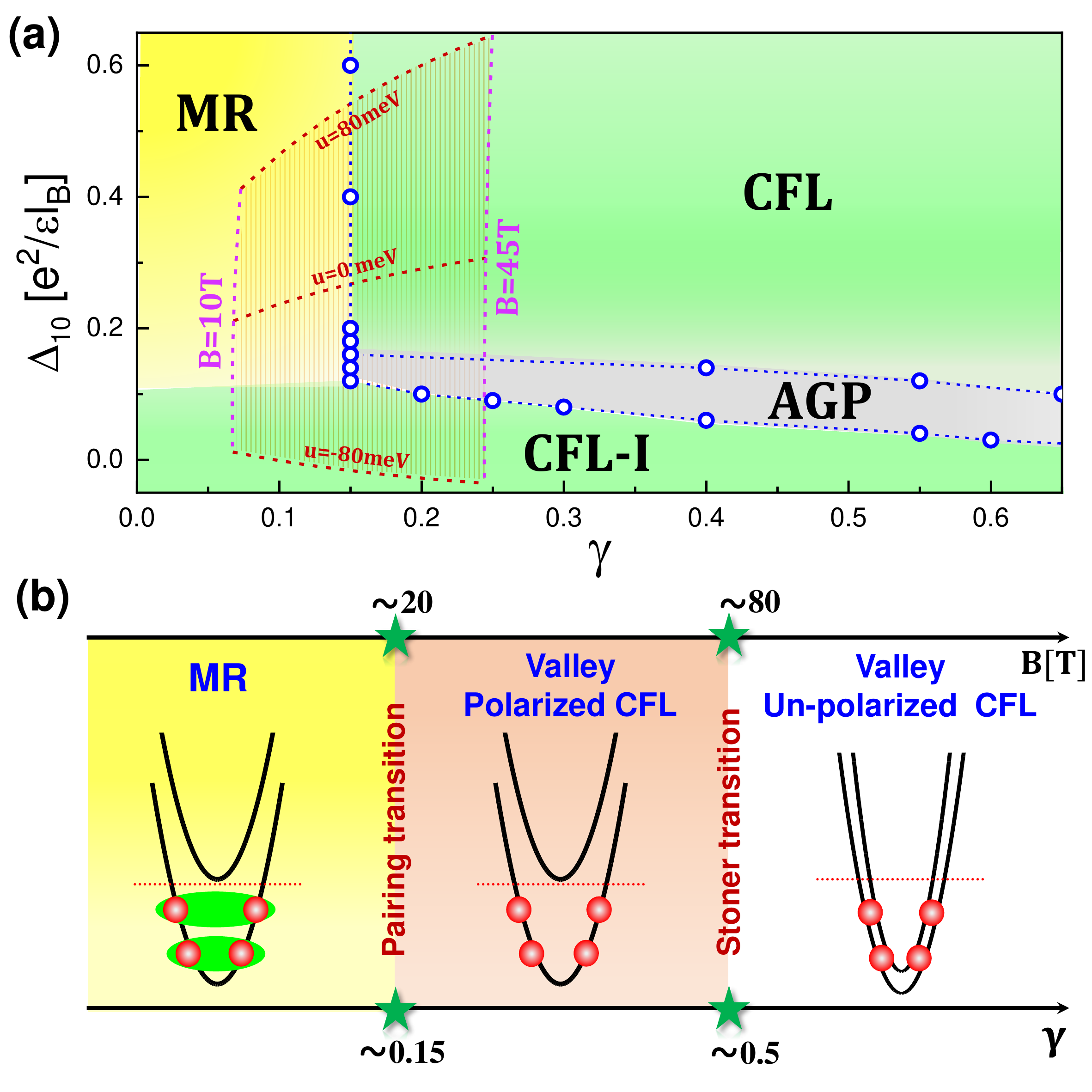}
\end{center}
\par
\renewcommand{\figurename}{Fig.}
\caption{(Color online) (a) Phase diagram of two-orbital model. $\Delta_{10}$ is the orbital splitting and $\gamma$ parametrizes form factors controlled by magnetic field. We identify the MR, two types of CFL states, and an intermediate anisotropic gapless phase (AGP).  The shaded region is the expected range of parameters accessed in BLG by tuning $B[T]$ and the interlayer electric bias $u$. (b) The phase diagram of SU(2) two-valley model [see Eq.~\ref{Eq:Ham}] as a function of $B[T]$ or $\gamma$ expected to be realised at $u= 0$. There are three phases: the valley polarized MR, and the valley polarized and un-polarized CFL states.
}
\label{Fig:PhaseDiagram}
\end{figure}

 {\textbf{Introduction.}} The advancements in the quality of graphene and the increased sophistication of techniques to probe it have positioned it as a rich platform to study the strongly
correlated physics of the quantum Hall regime. Recent hallmarks of this progress include the observation of bubble phases in monolayer graphene~\cite{Dean2018}, even denominator
fractional quantum Hall states near a pseudo-spin transition in monolayer graphene~\cite{Young2018}, fractional Chern insulators in graphene-hexagonal boron nitride
hetero-structures~\cite{Young2018}, even denominator fractional quantum Hall states in bilayer graphene~\cite{Morpurgo2014,Young2017,Dean2017}, the observation of exciton condensation in double
bilayer graphene~\cite{Kim2016,Dean2017b}, and new sequences of interlayer correlated fractional quantum Hall states in double-layer graphene~\cite{Dean2019}.

In this letter, we would like to offer a proposal to reap yet another fruit of this progress. We will show that bilayer graphene (BLG) is an ideal platform to realize a particularly clean quantum phase transition between two remarkable fractionalized phases of matter: the composite fermi liquid (CFL) metal~\cite{HLR} and the non-Abelian Moore-Read (MR) fractional quantum Hall state~\cite{MR}. Our study builds upon previous numerical studies~\cite{Apalkov,Papic2011,Papic2014,Young2017,Thomale2011} by incorporating our recently refined understanding of the Hamiltonian of the nearly eightfold degenerate zero Landau level of BLG~\cite{Young2017b}. There are two key ingredients that allow to controllably tune through this phase transition. One of them, first recognized in  Ref.~\cite{Apalkov}, is that the cyclotron orbital character of one of the Landau levels can be tuned continuously from mostly $n=1$ character at small perpendicular magnetic fields into mostly $n=0$ at high perpendicular fields. The second is the ability to enhance the splitting between $n=0$ and $n=1$ cyclotron orbits via the interlayer electric bias~\cite{Thierry2019}, whereby reducing the quantum fluctuations that make the MR
state unexpectedly strong at zero interlayer bias in experiments~\cite{Young2017,Dean2017} in order to facilitate its quantum melting into the CFL state. The expected phase diagram is depicted in Fig.~\ref{Fig:PhaseDiagram}.

Theoretically the MR state can be understood as a $p+ip$ paired state of the CFL~\cite{Greiter92,ReadGreen}. Unlike ordinary metals, the CFL has been argued to not have generic pairing instabilities at low temperatures~\cite{Bonesteel,Metlitski}, although an earlier study claimed the contrary~\cite{Greiter92}. If the CFL is stable against pairing, it would be possible to have an ideal stable phase transition from it into MR state by adding sufficiently large perturbations to the Hamiltonian. Originally, it was argued that this  transition would be first order~\cite{Bonesteel}, but this conclusion was challenged more recently by studies that argued that a stable continuous phase transition between the CFL and MR states is possible~\cite{Metlitski,Wang}. Numerical studies support a possible continuous transition~\cite{HR2000,Moller,Papic2012}, although a definitive numerical conclusion is currently out of reach due to system size limitations, we have found certain features in the finite size spectra that indicate a possible continuous phase transition~\cite{supplementary}.

Experimentally, the phase transition has been studied by tuning subband level crossings~\cite{Shayegan2011,Smet2012} and more recently hydrostatic pressure~\cite{Csathy2016,Csathy2017,Csathy2018} in GaAs quantum wells. The sub-band level crossing, however, produces a rather abrupt change of the microscopic parameters of the Hamiltonian and the transition is therefore likely first order~\cite{Papic2012}. The isotropic hydrostatic pressure experiments, found the MR state transitions into a compressible phase with anisotropic transport properties, in resemblance to the transitions induced by applying in-plane field~\cite{Pan1999,Lilly1999,Gabor2005,Zheng2017}, and therefore potentially placing the problem on a different universality class from that of interest here. Additionally, one limitation of the pressure-driven platform is that it is difficult to capture it with an ideal Hamiltonian. Further details on these precedents are in ~\cite{supplementary}.

\begin{figure}[btp]
\begin{center}
\includegraphics[width=0.4\textwidth]{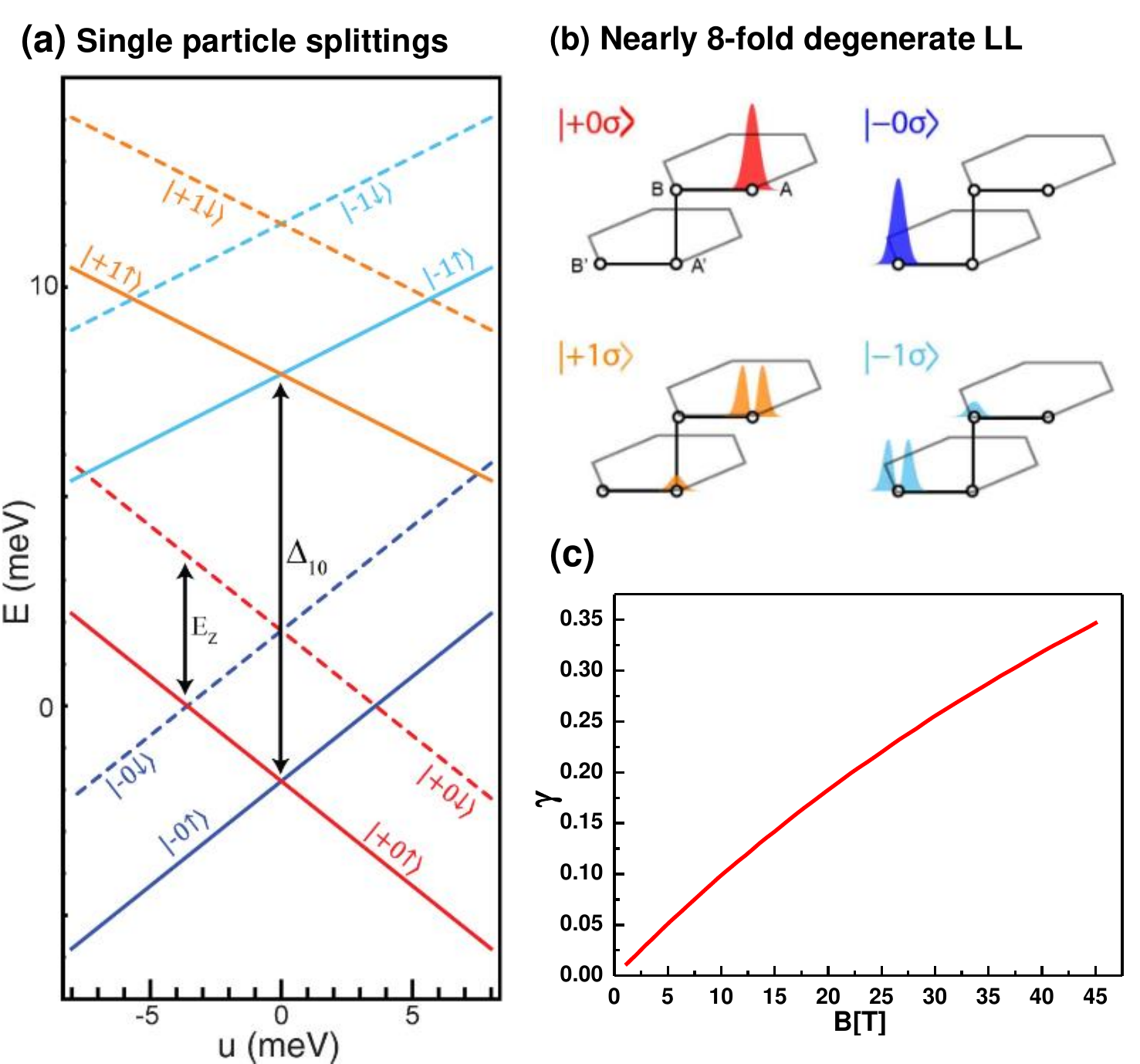}
\end{center}
\par
\renewcommand{\figurename}{Fig.}
\caption{(Color online) (a) The single particle splittings of BLG as a function of interlayer bias $u$. (b) The schematic depiction of the zero Landau level (ZLL) manifold of BLG $|\tau n \sigma \rangle$ on $(A,B',A',B)$ sites.  Here, $\tau=\{K,K'\}\equiv \{+,-\}$ denotes valleys, $n$ and  $\sigma$ are LL and spin index, respectively. (c) The relationship between parameter $\gamma$ and the magnetic field $B$. Figures (a) and (b) are from~\cite{Young2017b}, the data of (c) is from~\cite{Young2017}.}
\label{Fig2}
\end{figure}

{\textbf{Models and Key Results.}}  The zero Landau level (ZLL) manifold of BLG comprises eight internal Landau levels that we denote by $\psi_{n,\tau,\sigma}$, where $n=\{0,1\}$, $\tau=\{K,K'\}$  and $\sigma=\{\uparrow,\downarrow\}$ designates orbits, valleys and spins labels respectively.
The $\psi_1$ orbitals can be approximated as having weight on the $n=0$ and $n=1$  cyclotron Galilean orbitals (denoted by $\phi_{0,1}$) ~\cite{Young2017,Papic2014}: $\psi_{1,K}=(\sqrt{1-\gamma} \phi_{1},0,\sqrt{\gamma} \ \phi_{0},0)$ and $\psi_{1,K'}=(0,\sqrt{1-\gamma} \ \phi_{1},0,\sqrt{\gamma} \ \phi_{0})$.
Here, the different components denote amplitudes on $(A,B',A',B)$ sites in Fig.~\ref{Fig2} (b), and $\gamma \in [0,1]$ is a parameter controlled primarily by the perpendicular magnetic field whose typical values are shown in Fig.~\ref{Fig2} (c). On the other hand, the $\psi_0$ orbitals can be approximated as having only $n=0$ Galilean orbitals:~\cite{Young2017,Papic2014} $\psi_{0,K}=(\phi_{0},0,0,0)$ and $\psi_{0,K'}=(0, \phi_{0},0,0)$. A general interaction Hamiltonian projected onto a multi-flavor Landau level can be written as:
\be \label{Eq:Ham}
V=\sum_{{\mathbf{q \{\alpha\}}}}\frac{v({\mathbf {q}})}{2A}F_{\alpha_1\alpha_2}({\mathbf {q}})F_{\alpha_3\alpha_4}({\mathbf {-q}}):\rho_{\alpha_1\alpha_2}^\dagger({\bf q})\rho_{\alpha_3\alpha_4}({\bf q}):,~
\ee
where $v(\mathbf{q})$ is the Fourier transform of the un-projected interaction, $F_{\alpha \alpha'}({\mathbf q})$ is the density form factor determined by the wavefunctions, and $\rho_{\alpha \alpha'}({\mathbf q})$ are the flavor resolved intra-Landau-level guiding center density operators (see e.g.~\cite{Book}). We set magnetic length $l_B\equiv\sqrt{\hbar c/eB}$ as  the unit of length and $e^2/{\varepsilon l_B}$ as the unit of energy.

Our ideal Hamiltonian of interest is comprised of the Coulomb interaction projected onto the single $\psi_{1,\tau}$ Landau level. The form factor $F_{11}({\bf q})$ is given by
$F_{11}({\bf q}) = (1-\gamma) F_1({\bf q})+\gamma F_0({\bf q})$,
where $F_{0,1}({\bf q})=\exp(-{\bf q}^2/4) L_{0,1}[{\bf q}^2/2]$  are the form factors for $n=0$ and $n=1$ Galilean Landau levels. Therefore the Hamiltonian continuously interpolates from a
$n=1$  Galilean Landau level at small $\gamma$ (weak perpendicular fields) to an $n=0$ Galilean Landau level at large $\gamma$ (strong perpendicular fields).
We have found that at half filling the MR is the ground state of this ideal Hamiltonian for $\gamma \lesssim 0.15$ whereas for $\gamma\gtrsim 0.15$ the CFL is the ground state. To demonstrate that this conclusion remains robust in the presence of other flavors and to delineate the region of parameters to realize such ideal limit within more realistic models,
we will study several modifications to this ideal Hamiltonian.

The first modified Hamiltonian is an SU(2) symmetric version of the ideal Hamiltonian we just described, containing two-valleys $\psi_{1\alpha}$,  $\alpha=\{K,K'\}$.  Therefore the form factors are $F_{\alpha,\alpha'}({\bf q})=\delta_{\alpha,\alpha'}  F_{11}({\bf q})$.
 In this case we will show that the ground state spontaneously polarizes onto a single valley for $\gamma \lesssim 0.5$ and therefore the phase transition region from MR to CFL remains unmodified by the presence of a second degenerate valley (or spin). This interesting regime of vanishing single-particle valley splitting with spontaneous valley polarization can be best achieved at total filling $\nu=3+1/2$ near zero interlayer bias $u\approx 0$. In the supplementary we describe how valley dependent interactions, which break the SU(2) valley symmetry down to U(1), are not expected to significantly affect the location of the phase transition.

\begin{figure}[!t]
\begin{center}
\includegraphics[width=0.5\textwidth]{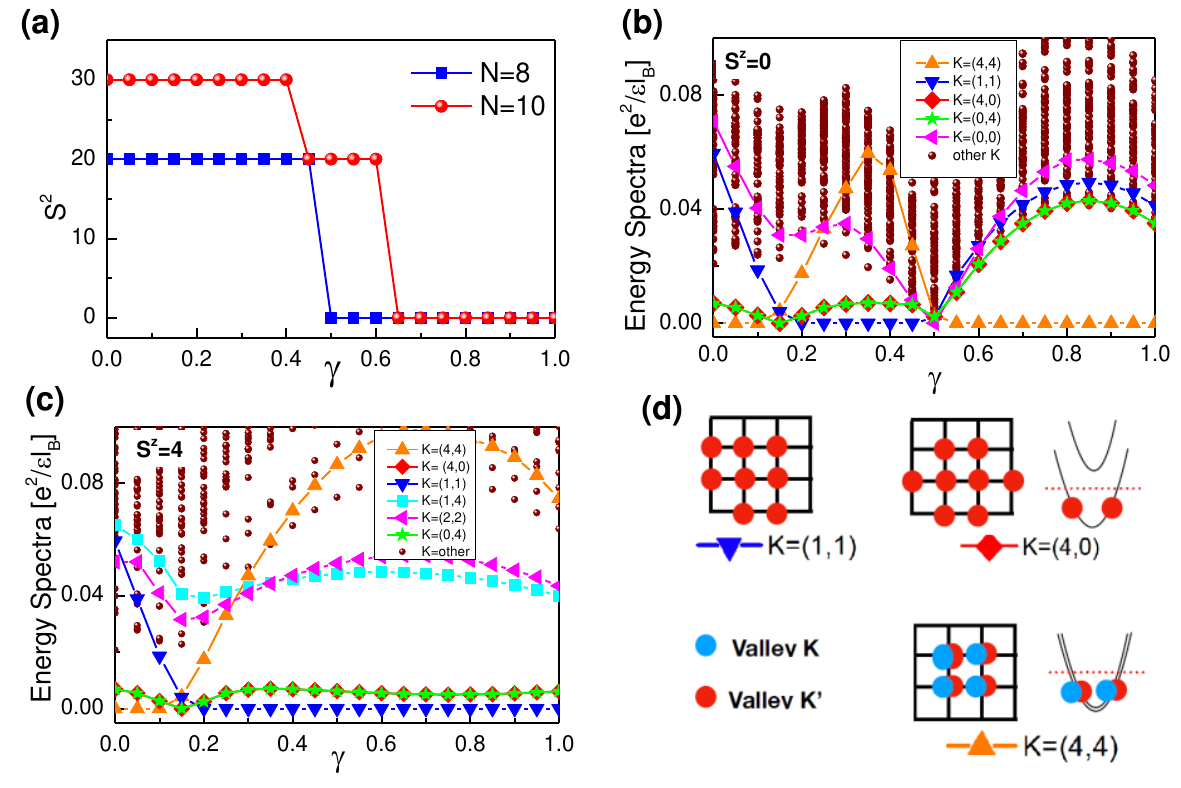}
\end{center}
\par
\renewcommand{\figurename}{Fig.}
\caption{(Color online) (a) The SU(2) Casimir operator $S^2$ as a function of $\gamma$.   (b) and (c) show the energy spectra as a function of $\gamma$ for valley-unpolarized sector (b) and valley-polarized sector (c). (d) The cluster of momentum determined by the trial wave function for polarized CFL (upper) and unpolarized CFL (lower).}
\label{Fig3}
\end{figure}

The second modified Hamiltonian contains two Landau levels with different orbital character and the same valley $\psi_{\alpha K}$, $\alpha =\{0,1\}$. There is no flavor conservation in this case and thus the form factors contain flavor off-diagonal components, and are given by: $F_{00}({\bf q}) =F_{0}({\bf q})$,  $F_{11}({\bf q}) =(1-\gamma) F_1({\bf q})+\gamma F_0({\bf q})$, $F_{10}({\bf q})=[F_{01}(-{\bf q})]^*= \sqrt{1-\gamma} \exp(-{\bf q}^2/4)[-i(q_x+iq_y)/\sqrt{2}]$.
We also add a single particle spliting $\Delta_{\mathrm{10}}$ between these two orbital flavors.
For this model, we will demonstrate that a single
particle splitting, $\Delta_{10}$, favouring $\psi_{1,K}$ over $\psi_{0,K}$ of about $\Delta_{10} \gtrsim 0.2 e^2/\epsilon l_B$, is enough to reach the behavior of the ideal Hamiltonian containing only the  $\psi_{1,K}$ Hamiltonian previously described. According to our estimates the built-in splitting between these orbitals in BLG is sufficient to reach this limit, but additionally, we will show that this splitting can be further enhanced at total filling $\nu=-3+1/2$ by applying interlayer effective field, giving us confidence that the ideal regime to realize the single-component MR to CFL transition can be accessed in BLG, in agreement with previous DMRG studies~\cite{Young2017}.  

We will resort to numerical exact diagonalisation in the torus geometry \cite{Haldane1985,Yoshioka1984} to investigate the nature of the ground states as a function of $\gamma$ and $\Delta_{10}$. Throughout the main body of this paper we will focus on the Coulomb interaction, $v({\bf q})=2\pi e^2/\epsilon |{\bf q}|$, however, in the supplementary we demonstrate that the key conclusions remain for more realistic interactions that account for screening~\cite{supplementary}.

{\textbf{Two-valley model.}} We begin by studying  an SU(2) symmetric model including the $\psi_{1K}$ and $\psi_{1K'}$ valleys. The ideal Hamiltonian describing a single valley can be obtained simply by restricting to the SU(2) subspace with maximal valley polarization.
 We denote the valley polarization as $S_z=(N_{K}-N_{K'})/2$. Figure~\ref{Fig3} (a) depicts the value of SU(2) Casimir operator,  $S^2$, which determines the valley polarization of the ground state as a function of $\gamma$. We find that the system jumps from a polarized state into a singlet at $\gamma\sim 0.5$, although a small intermediate range of $\gamma$ with partial polarization cannot be completely discarded. It is well documented, experimentally~\cite{Du95,Kukushkin99,Tiemann2012} and numerically~\cite{Park1998,Park2001,Zheng2018}, that in the SU(2) limit the CFL in the $n=0$ LL ($\gamma\rightarrow 1$) is a two-component unpolarized singlet. It is also well-established that in the SU(2) limit of an $n=1$LL ($\gamma\rightarrow 0$) the MR state is a fully polarised ferromagnet spontaneously breaking the SU(2) symmetry~\cite{Morf,Park1998,Feiguin}. The polarization we find is consistent with these expectations and it is therefore natural to conclude that in these limits we have a valley singlet CFL state at $\gamma\rightarrow 1$ and a valley polarized state at $\gamma\rightarrow 0$. However, we have found another phase at intermediate $\gamma$, namely, a single component Stoner-type CFL with spontaneous valley polarization.

We will now show that the quantum numbers of the states for $0.15 \lesssim \gamma\lesssim 0.5$ indeed are those of a fully polarized CFL while those of the state present for $\gamma \gtrsim 0.5$ correspond to a two-component un-polarized CFL. To do so, we consider trial CFL wavefunctions~\cite{RR1994} in the torus~\cite{Haldane1985,HR2000,Zheng2018,Read1994}.  We review the construction of these trial CFL wavefunctions in~\cite{supplementary}.
The key quantum number that allows direct comparison with numerics is the many-body momentum, which, for the square torus reads as ${\bf K}= (L/N)\sum_{i}{\bf k}_i=(2\pi/N) \sum_{i} (-m_{2i},m_{1i}) \ {\rm mod}(N)$ ($m_{1,2}\in Z \ {\rm mod} (N_\phi)$). This momentum in units of $(2\pi/N)$ is the same that labels the states of the spectrum in Fig.~\ref{Fig3} (b) and (c)~\cite{footnote1}. The cluster of momentum that correspond to the states that minimise the trial mean-field energy of a single component CFL~\cite{Zheng2018} are $(K_x,K_y)\in (2\pi/N) \{(1,1),(4,0),(0,4)\}$ for $N=8$ particles and are shown in Fig.~\ref{Fig3} (d). We see that these states have the same quantum numbers of those obtained from exact diagonalisation for $0.15 \lesssim \gamma\lesssim 0.5$. Following a similar analysis for a two-component CFL singlet state~\cite{footnote2}, one can show that for $8$ particles there is a unique finite size cluster forming a {\it closed shell} in momentum, namely, that the lattice of displacement vectors transforms trivially under the point group of the square torus. This state forms at momentum $(K_x,K_y) = (2\pi/N) (4,4)=(\pi,\pi)$ and is depicted in Fig.~\ref{Fig3} (d). This is indeed coincides with the momentum of the ground state realized for $\gamma \gtrsim 0.5$ in Fig~\ref{Fig3} (b) .

Therefore, we have found that the SU(2)-valley invariant system has three phases: (I) a valley polarized MR Pfaffian state for $\gamma\lesssim 0.15$, (II) a valley polarized single component CFL state for $0.15 \lesssim \gamma\lesssim 0.5$, and (III) a valley un-polarized two component CFL state for $\gamma\gtrsim 0.5$, as illustrated in Fig.~\ref{Fig:PhaseDiagram} (b). For the ideal Hamiltonian from Eq.~\eqref{Eq:Ham} of a single valley case then we would simply encounter phases (I) and (II). In the supplementary material~\cite{supplementary} we demonstrate that these conclusions still hold for larger system sizes and for more realistic screened versions of the Coulomb interaction, and we also provide arguments for why the essential physics of the ground states under consideration are robust.
For bilayer graphene, this leads us to expect that the transition between MR and CFL can be achieved near $\nu=\pm 3+1/2$ for $B \sim 20T$ [see Fig.~\ref{Fig:PhaseDiagram} and Fig.~\ref{Fig2}(b) ]. Unfortunately, the Stoner type transition between the single and two-component CFL states is expected at about  $B \sim 80T$. A related Stoner transition between single and two component CFL states has been recently discussed in monolayer graphene~\cite{Jain2015}.

\begin{figure}[!t]
\begin{center}
\includegraphics[width=0.5\textwidth]{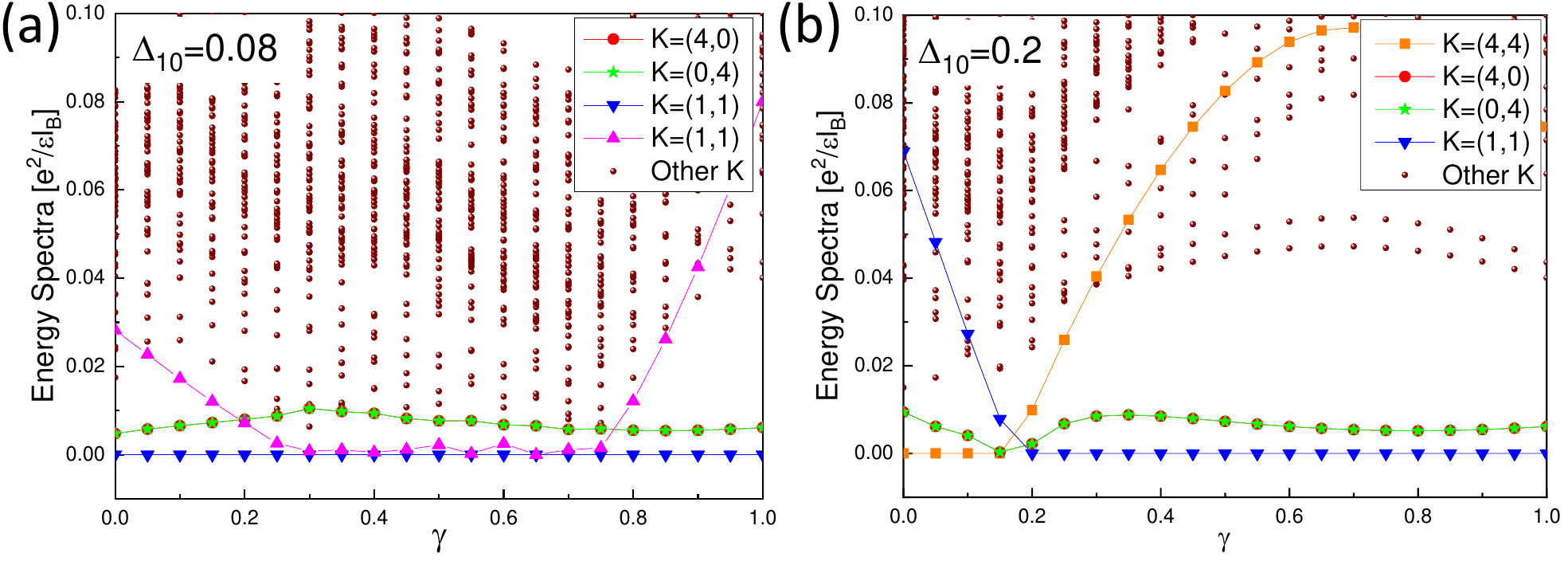}
\end{center}
\par
\renewcommand{\figurename}{Fig.}
\caption{(Color online)The energy spectra of two-orbital model as a function of $\gamma$ at different single particle splitting $\Delta_{10}$ between $\psi_0$ and $\psi_1$ orbits: (a) $\Delta_{10}=0.08$, (b) $\Delta_{10}=0.2$. }
\label{Fig4}
\end{figure}

{\textbf{Two-orbital model.}} A special feature of the zeroth Landau level of BLG is the relatively small energy splitting between the $\psi_1$ orbits with $n=1$ cyclotron character and the $\psi_0$ orbits with $n=0$ cyclotron character. Therefore it is important to assess how robust the phases are to quantum fluctuations between these levels. To do so, we consider a model with the Coulomb interaction, $v({\bf q})=2\pi e^2/\epsilon |{\bf q}|$, projected onto these two levels and an additional single particle splitting $\Delta_{10}$ favouring $\psi_1$.  At total filling $\nu=1/2$ it is clear that in the limit in which $\Delta_{10}\gg e^2/{\epsilon l_B}$, we will recover the physics of the ideal limit containing only the half-filled $\psi_1$ orbits.

The energy spectra of two-orbital model versus $\gamma$ and $\Delta_{10}$ are shown in Figs.~\ref{Fig4} and  supplementary material~\cite{supplementary}. We have found that this limit is achieved by a splitting $\Delta_{10} \gtrsim 0.2 e^2/{\epsilon l_B}$ as shown in Fig.~\ref{Fig:PhaseDiagram}(a) and Fig.~\ref{Fig4} (b). At smaller $\Delta_{10}$ we have found another CFL state  labelled CFL-I  in Fig.~\ref{Fig:PhaseDiagram} (a).
 CFL-I is the ordinary CFL realized at half-filled $\psi_0$ orbit. The reason why this CFL becomes the ground state near $\Delta_{10}=0$ is that there is an exchange energy gain to occupy $n=0$ orbits due to the smaller spatial extension and hence larger exchange holes~\cite{Book}, and therefore at half-filling the state is the conventional $n=0$ CFL.
Here, $\Delta_{10}$  splits the degeneracy of the MR Pfaffian and anti-Pfaffian. An interesting possibility is that it could be possible, by controlling the interlayer bias, to tune experimentally from Pfaffian to anti-Pfaffian by changing the sign of $\Delta_{10}$, as further discussed in \cite{supplementary}.

Interestingly, we have also encountered an anisotropic gapless phase (AGP) at intermediate orbital splitting and magnetic field in  Fig.~\ref{Fig:PhaseDiagram}(a).
This phase features a multiplicity of low lying states and a robust ground state quasi-degeneracy indicative of a gapless broken symmetry state, as shown in Fig.~\ref{Fig4} (a) and in the supplementary. Additionally the spectrum has a high sensitivity to changes of the aspect ratio of the torus, which indicates the breaking of rotational symmetry, shown in the supplementary~\cite{supplementary}. This phase could be accessed in BLG near filling $\nu=3+1/2$ and therefore we hope that future numerical and experimental studies can shed more light on its nature.
We find that many phases meet near the $n=1/0$ level crossing around $\Delta_{10}\sim 0.11 e^2/\epsilon l_B$ and $\gamma \lesssim 0.15$~\cite{supplementary}. It is possible that other phases might be stabilized near this level crossing, such as the non-Abelian $221$ parton state, as advocated in Ref.~\cite{Wu2017}.We hope that future studies will further address this interesting possibility.

{\textbf{Region of parameters accessed in BLG}}. In Fig.~\ref{Fig:PhaseDiagram}(a) we have superimposed the expected range of parameters~\cite{NoteExp} that can be accessed in BLG by tuning perpendicular magnetic field and the interlayer electric bias $u$.  
The region for $u \leq 0$ should be accessible at $\nu=3+1/2$, while the region of positive $u\geq 0$ should be accessible for $\nu=-3+1/2$. This can be inferred from Fig.\ref{Fig2} which shows that the single particle level splitting $\Delta_{10}$ decreases with $|u|$ for $\nu=3+1/2$ and increases with $|u|$ $\nu=-3+1/2$. Therefore, $\nu=-3+1/2$ can be brought much closer to the ideal limit to study the ideal CFL to MR transition by applying large interlayer bias, although different physics could be potentially accessed $\nu=3+1/2$ with the interlayer bias, such as the nearly SU(2) valley symmetric conditions for $u\approx 0$  and the new intermediate anisotropic gapless phase (AGP) shown in Fig.~\ref{Fig:PhaseDiagram}(a).

{\textbf{Discussion and summary.}} We have advanced a proposal for realising a particularly clean and widely tuneable phase transition between the MR and CFL states in BLG at fillings $\nu=\pm 3+1/2$. The phase transition can be tuned by the perpendicular magnetic field, as in phase transitions previously realised in monolayer graphene~\cite{Yacoby1,Yacoby2,Young2018,Thomale2011}. The simplest version of this phase transition is better achieved at $\nu=-3+1/2$ at large interlayer biases $|u|\gtrsim 50meV$, where we have demonstrated that both valley and orbital fluctuations become insignificant. This filling factor at such interlayer biases is therefore an ideal platform to study the CFL to MR transition. At the filling factor $\nu=3+1/2$ one encounters increased valley fluctuations for vanishing interlayer bias, where one expects a near SU(2) valley symmetry. We have shown that this symmetry is spontaneously broken and the system is also expected to transition from a spontaneously valley polarised MR state into a spontaneously valley polarised Stoner CFL enriched by the physics of valley symmetry breaking. At this filling the interlayer electric field tends to enhance $n=0/1$ orbital fluctuations, and this can be used to access a potentially new anisotropic gapless phase (AGP) near the level crossing of $n=0$ and $n=1$ orbits of a common valley as shown in Fig.~\ref{Fig:PhaseDiagram}(a).

\begin{acknowledgments}
We thank Liang Fu, T. Senthil, Jainendra Jain, Andrea Young and Ashvin Vishwanath for valuable discussions.
This work  was supported by the U.S. Department of Energy (DOE), Office of
Basic Energy Sciences under Grant No. DEFG02-06ER46305. Z.Z. would like to gratefully acknowledge the computational resources at
California State University Northridge for performing the numerical simulations.
\end{acknowledgments}

\begin{center}
\textbf{\large Supplemental Material}
\end{center}

\renewcommand{\theequation}{S\arabic{equation}}
\setcounter{equation}{0}
\renewcommand{\thefigure}{S\arabic{figure}}
\setcounter{figure}{0}
\renewcommand{\bibnumfmt}[1]{[S#1]}

\section{Comments on theoretical and experimental precedents for the MR to CFL transition.}
 Theoretically the MR state can be understood as a $p+ip$ paired state of composite fermions~\cite{Greiter92s,ReadGreens}. Unlike ordinary electron fermi liquids, and contrary to earlier arguments~\cite{Greiter92s}, the CFL state is believed not to have generic pairing instabilities at low temperatures~\cite{Bonesteels,Metlitskis} but a quantum
phase transition could be driven by adding a sufficiently large perturbation to the Hamiltonian.
An early theoretical study of this transition concluded that gauge fluctuations would drive the transition generically first order~\cite{Bonesteels}, but this conclusion has been challenged by more recent studies that concluded that a stable continuous phase transition between the CFL and MR states is possible~\cite{Metlitskis,Wangs}. On the other hand, numerical studies have found the transition to appear continuous~\cite{HR2000s,Mollers,Papic2012s}. Therefore, experimental studies could shed fundamentally new light on the very nature of this transition.

Experimentally the phase transition has been studied by tuning subband level crossings~\cite{Shayegan2011s,Smet2012s} and more recently by applying hydrostatic pressure~\cite{Csathy2016s,Csathy2017s,Csathy2018s} to 2DEGs in GaAs quantum wells. Studies in which the MR state is destroyed by in-plane fields, although interesting in their own right, do not probe the ideal universality of the continuous phase transition between MR and CFL that we are considering as they explicitly break the isotropy of the problem. The sub-band level crossing had the interesting effect of enhancing the gap of the MR state as the crossing is approached~\cite{Shayegan2011s,Smet2012s,Papic2012s}, however, the crossing produces a rather abrupt change of the microscopic parameters of the Hamiltonian and the transition is therefore highly likely first order as illustrated by the sharp collapse of the wavefunction overlap in numerical studies~\cite{Papic2012s}.

The hydrostatic pressure experiments, on the other hand, revealed an interesting but intriguing scenario. As the pressure is tuned trough a critical value $P_{c1}$, the MR state transitions into a compressible phase with anisotropic transport properties, which in turn transitions at a larger pressure $P_{c2}$ into an isotropic compressible phase. Experiments have been argued to be consistent with a continuous phase transition between the MR and the anisotropic phase~\cite{Csathy2017s}. However,  if indeed in an ideal clean limit this corresponds to a stable continuous phase transition between these phases of matter, it should lie beyond any of the paradigms that we understand so far. This is because, to the best of our knowledge, currently there exists no known theoretical scenario in which a MR state could have a clean stable critical point with any known anisotropic compressible phase. Even, if this anisotropic phase was a Pomeranchuk analogue of the CFL state~\cite{Fradkin2016s}, a putative critical point would appear to require the coincidence of two distinct events, namely, the vanishing of the composite fermion pairing gap and the breaking of the rotational symmetry of the parent CFL state, making it a fine tuned scenario. On the other hand, it is known from numerical studies that the MR state is in close energetic competition with anisotropic stripe phases~\cite{Rezayi1999s,HR2000s,Rezayi2000s,Zheng2017s}, and that the transition is likely first order~\cite{HR2000s}. Since first order phase transitions in two-dimensions can be driven second order by disorder~\cite{ImryMas,ImryWortiss,Aizenmans}, a natural possibility is that the phase transition driven by pressure in GaAs samples is a first order transition smeared by disorder effects. Additionally, from the numerical point of view, one limitation of the pressure-driven platform is that it is difficult to capture it with an ideal Hamiltonian: the pressure is believed to primarily change the Landau level mixing parameter of the 2DEG (and its effective width~\cite{Csathy2018s}). But, capturing the Landau level mixing reliably in numerical studies is a fairly difficult task.

Based on these precedents, we believe that finding a suitable platform to realise a clean phase transition between MR and CFL states is a much needed endeavour.
As we will argue BLG offers an unprecented platform to investigate this physics, because it can be tuned smoothly through this transition by tuning the perpendicular magnetic field, and because the microscopic Hamiltonian as a function of these parameters has been recently established in great detail.
We also would like to note in passing that recent studies have revealed the possibility of interesting new phases at a level crossing between $n=1$ and $n=0$ Landau levels at total effective filling $\nu=1/2$ in BLG~\cite{Young2017s,Maissams,Zaletel2018s} and in ZnO~\cite{Falsons}. Our proposed scenario differs from these in that there is no abrupt level crossing but rather a smooth transformation of the character of the landau level from $n=1$ to $n=0$ as a function of tuning parameters.

\begin{figure}[tbp]
\begin{center}
\includegraphics[width=0.5\textwidth]{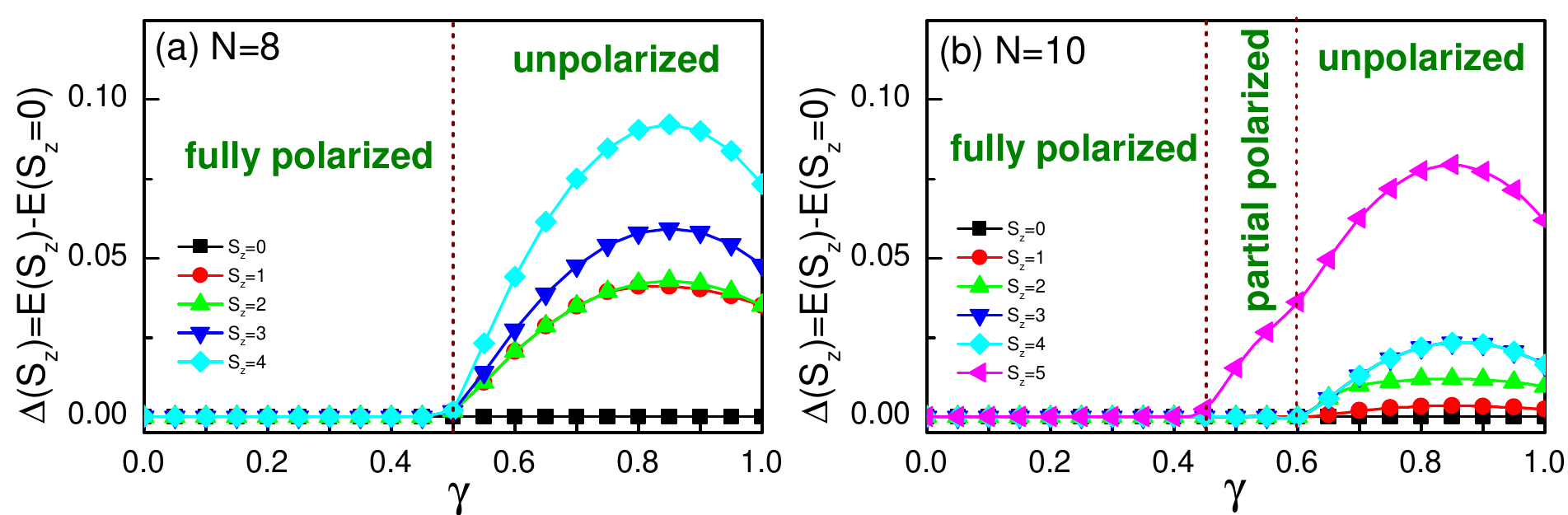}
\end{center}
\par
\renewcommand{\figurename}{Fig.}
\caption{(Color online) The energy difference $\Delta (S_z)$  between $S_z\geq0$ sector and $S_z=0$ sector as a function of $\gamma$  for $N_e=8$ (a) and $N_e=10$ (b)  systems. }
\label{Fig:Sz}
\end{figure}

\begin{figure}[tbp]
\begin{center}
\includegraphics[width=0.5\textwidth]{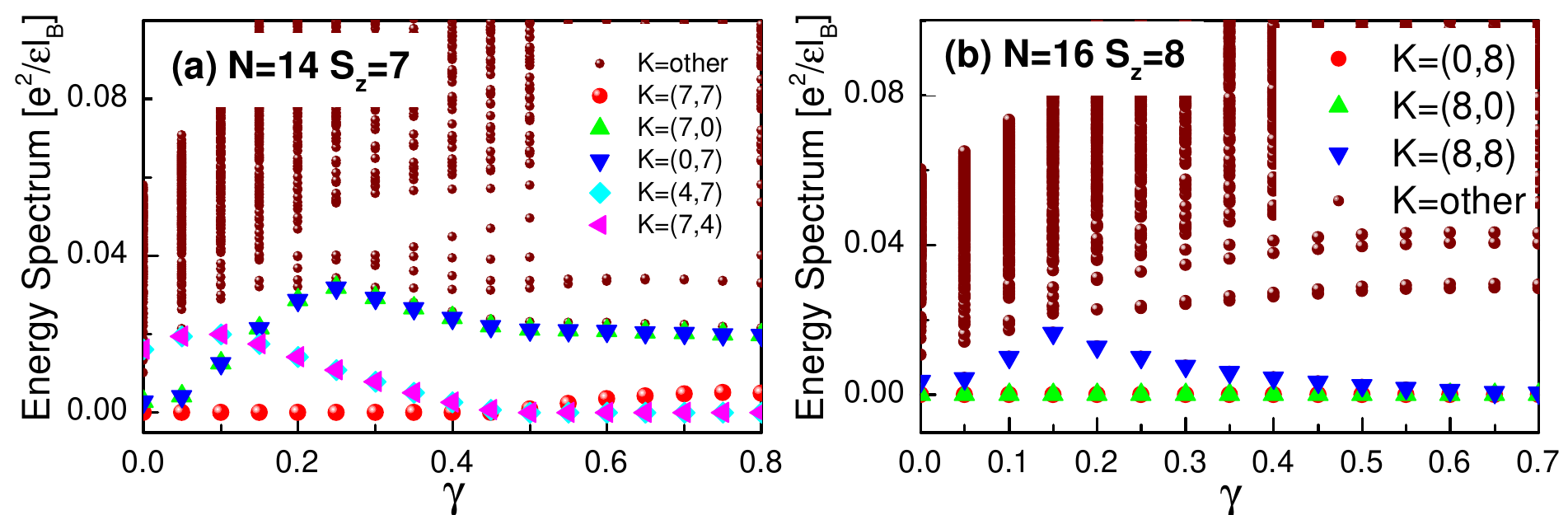}
\end{center}
\par
\renewcommand{\figurename}{Fig.}
\caption{(Color online) The energy spectra as a function of  $\gamma$: (a)$S_z=7$ sector for $N_e=14$ system, (b)$S_z=8$ sector for $N_e=16$ system.  }
\label{Fig:N14N16}
\end{figure}

\section{Two-valley models with SU(2) symmetry}\label{Sec:Two-valley}
We consider an SU(2) symmetric model including the  $\psi_{1K}$ and $\psi_{1K'}$ valleys and  denote the valley polarization as $S_z=(N_{K}-N_{K'})/2$. The $\psi_{1}$ orbits can be written as
\be\label{psi1}
\psi_{1,K}=
\begin{pmatrix}
\sqrt{1-\gamma} \ \phi_{1} \\
0\\
\sqrt{\gamma} \ \phi_{0}\\
0
\end{pmatrix}, \ \
\psi_{1,K'}=
\begin{pmatrix}
0 \\
\sqrt{1-\gamma} \ \phi_{1}\\
0 \\
\sqrt{\gamma} \ \phi_{0}
\end{pmatrix}.
\ee
The single valley can be realized in the SU(2) subspace with maximal valley polarizations $S_z=\pm N/2$. In the present exact diagonalization (ED)  simulation, we choose the square torus geometry with  length vectors $\mathbf{L_x}$ and $\mathbf{L_y}$. The projected Coulomb interaction given by 
\be
V=\frac{1}{2 A} \sum_{{\mathbf {q \{\alpha\}}}}v ({\bf q}) |F_{11}({\bf q})|^2 : \rho^\dagger({\bf q}) \rho({\bf q}):  ,
\ee
here $\alpha=\{K,K'\}$ denote valleys, the form factor $F({\bf  q})$ is
\be \label{Fq}
F_{11}({\bf q}) = (1-\gamma) F_1({\bf q})+\gamma F_0({\bf q}).
\ee
where $F_{0,1}({\bf q})=\exp(-{\bf q}^2/4) L_{0,1}[{\bf q}^2/2]$ are the form factors for $n=0$ and $n=1$ Galilean Landau levels (LLs). Within the Landau gauge,  the number of Landau orbits $N_\phi$ are determined by the area of the torus based on $|\mathbf{L_x}\times \mathbf{L_y}|=2\pi N_\phi$. We have set the magnetic length $l\equiv\sqrt{\hbar c/eB}$ as the unit of the length.

To identify the nature of  CFL, we consider trial CFL wavefunctions in the torus. The use of these wavefunctions to understand the quantum numbers and the shell filling effect of two-component CFL states in the torus has been recently discussed in Ref.~\onlinecite{Zheng2018s}, and we summarize the main points here. Single particle translations in a torus belong to a discrete lattice:
${\bf d} \in (m_1 {\bf L_1}+m_2 {\bf L_2})/N_\phi, \, m_{1,2}\in \mathbb{Z}{\rm mod}(N_\phi)$, where ${\bf L_{1,2}}$ are the principal vectors of the torus. To write down a trial single-component CFL  state with $N$ particles we draw a set of $N$ distinct vectors from this lattice $\{{\bf d}_i\}$. The trial state reads as:
\be\label{CFL}
|\Psi_{\text{CFL}}(\{{\bf d}_i\})\rangle=\det (\hat{t}_j({\bf d}_i)) | \Phi^{\text{Bose}}_{1/2}\rangle.
\ee
\noindent Here $| \Phi^{\text{Bose}}_{1/2}\rangle$ is the bosonic Laughlin wavefunction at $\nu=1/2$, $\hat{t}_j({\bf d})$ is the magnetic translation operator acting on particle $j$ by an amount ${\bf d}$. The displacement vectors can be associated with a dipole moment which in turn is related to the composite fermion momentum as ${\bf k}_i\equiv  l^{-2} \hat{{\bf z}}\times {\bf d}_i$~\cite{Zheng2018s}. Therefore the set of translation vectors $\{{\bf d}_i\}$ is a discrete and rotated version of the composite fermion fermi surface. The energetics of the state is such that the vectors tend minimise their difference so that, heuristically, an emergent kinetic energy can be argued to have the form:
\be\label{Etrial}
E[\{{\bf k}_i\}] \approx \ E_0+ \sum_{i<j} |{\bf k}_i-{\bf k}_j|^2/(2 m^* N).
\ee
\noindent The key quantum number that allows direct comparison with numerics is the many-body momentum, which, for the square torus reads as ${\bf K}= (L/N)\sum_{i}{\bf k}_i=(2\pi/N) \sum_{i} (-m_{2i},m_{1i}) \ {\rm mod}(N)$. This momentum in units of $(2\pi/N)$ is the same that labels the states of the spectrum. We only show states in the upper quarter of the many-body Brillouin zone since the remainder are related by point group symmetry.

 \textbf{Stoner transition.}---We begin with valley polarization process as a function of $\gamma$  based on Eq.~\ref{Fq}.  In the main text, we determine the valley polarization by examining the Casimir operator $S^2$. Here we target different valley polarization sector $S_z$ to compare the energy difference $\Delta (S_z)$ defined by $\Delta (S_z)\equiv E_0(S_z)-E_0(S_z=0)$,   to determine valley polarization $S^z$ of the ground state . $\Delta (S_z)$ represents the energy difference between the lowest energy in each  $S^z\geq0$ sector and  the lowest energy in $S^z=0$ sector. Figure~\ref{Fig:Sz} (a) and (b) show $\Delta (S_z)$ as a function of  $\gamma$ for both $N_e=8$  and $N_e=10$ systems. When $\gamma \lesssim 0.5$, the ground states of the systems are fully polarized with $S^z=N/2$, which is almost independent on the system size. The system  translates into the unploarized state when $\gamma \gtrsim 0.5$ for $N=8$ system, while there is an intermediate partial polarized phase ($0.5\lesssim \gamma \lesssim 0.6$) for $N_e=10$ system before the unploarized phase at $\gamma \gtrsim 0.6$. Such intermediate phase is favored when $N_e=10$ due to the shell filling effect of placing CFL on finite sized torus. For both fully polarized CFL and un-polarized CFL, $N_e=8$ system always has completely filled shells, which means that shell filling is not biasing towards either state at this system size.

\begin{figure}[tbp ]
\begin{center}
\includegraphics[width=0.5\textwidth]{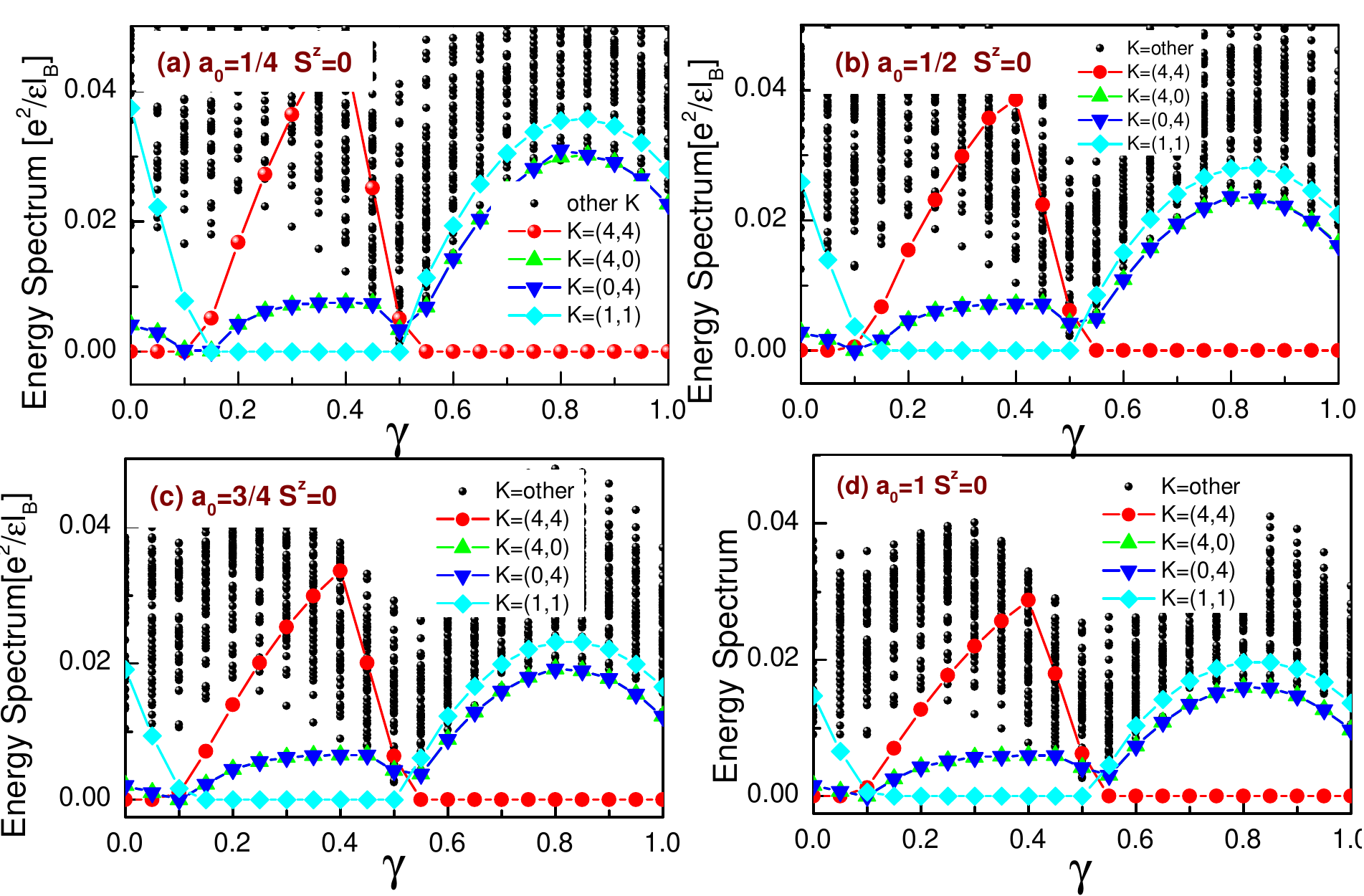}
\end{center}
\par
\renewcommand{\figurename}{Fig.}
\caption{(Color online) The energy spectrum as a function of $\gamma$ for $a_0=1/4$ (a), $a_0=1/2$ (b),$a_0=3/4$ (c),$a_0=1$ (d). }
\label{Fig:Screen}
\end{figure}

 \textbf{Pairing transition.}---When $\gamma=0$,  the form factor in Eq.~\ref{Fq} reduces to $F_{11}({\bf q})=\exp(-{\bf q}^2/4) L_1[ {\bf q}^2/2]$,  Fig.~\ref{Fig:Sz} shows that the ground state energy has perfectly flat dependence on valley polarization $S_z$, suggesting the fact that we have not only separate conservation of $S_z$ but also full $SU(2)$ invariance.  Thus the ground state is  a fully polarized magnet, which is simply the $2S_z+1$ symmetry related copies of the Pfaffian that are simply globally rotated by $SU(2)$ operations. The nature of the Pfaffian state can be identified by the  three-fold degeneracy on top of two-fold center of mass degeneracy in momentum sector $(K_x,K_y)/(2\pi/N)=(N/2,N/2),(0,N/2),(N/2,0)$.

 Figure~3 (b) and (c) in the main text show the low-lying energy spectra as a function of $\gamma$  in valley-un-polarized sector ($S_z=0$) and valley polarized sector ($S_z=N/2$ ), respectively.  When $\gamma \lesssim 0.5$, the ground state has $S_z=N/2$  so these two $S_z$ sectors display similar low-energy spectra. The topological sectors of MR state dominate the ground state property until $\gamma \sim 0.15$, where a phase transition takes place signaled by a level crossing in the spectra. This suggests two different phases in the polarized regime. Since the ground state is valley singlet for $\gamma \gtrsim 0.5$ , the phase survives only at   $0.15\lesssim \gamma \lesssim 0.5$. By comparing with the shell-filling calculation, the ground state is identified as the fully polarized CFL state [see Fig. 3 in the main text]. To check the finite size effect we calculate systems with several sizes, it turns out that the transition between MR and fully polarized CFL state are robust, as shown in the Fig.~\ref{Fig:N14N16} for $N=14$ and $N=16$ systems, the transition always happens near $\gamma \sim 0.15$, which is independent on system size.

\begin{figure*}[tbp ]
\begin{center}
\includegraphics[width=0.95\textwidth]{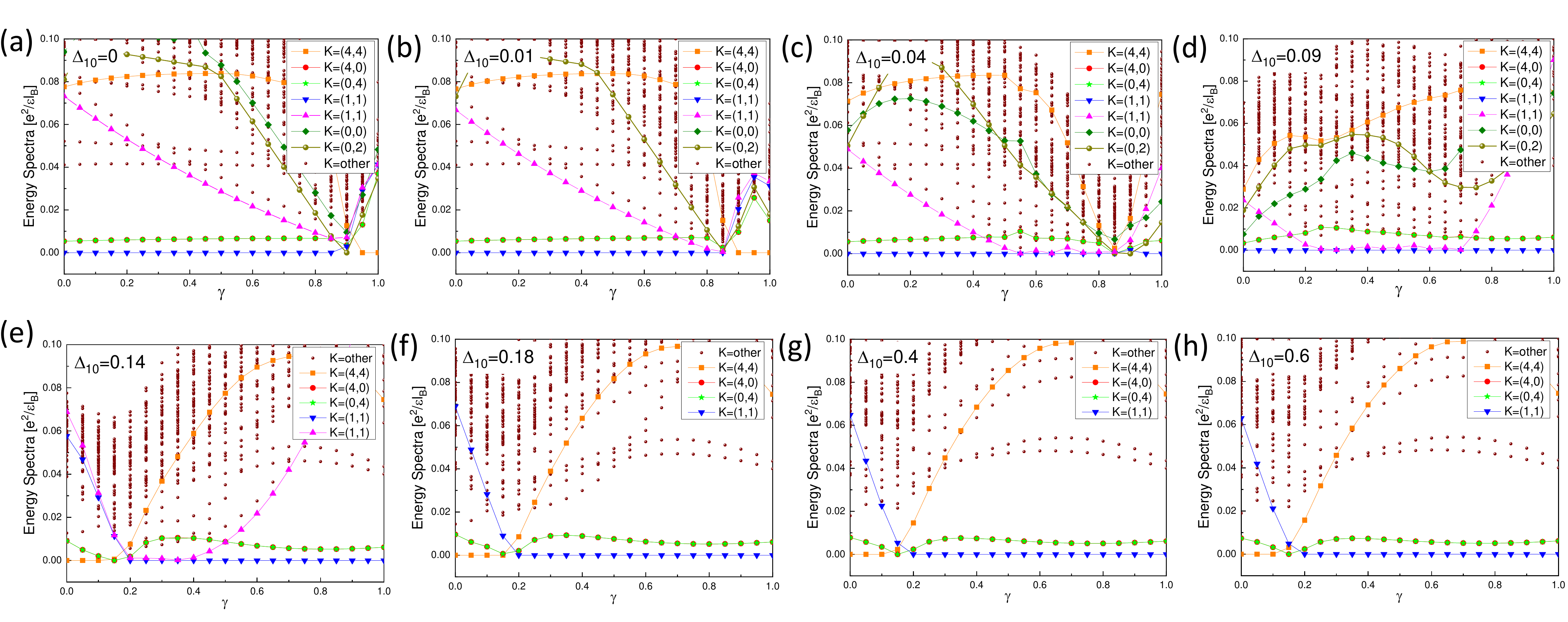}
\end{center}
\par
\renewcommand{\figurename}{Fig.}
\caption{(Color online) The energy spectrum as a function of $\gamma$ for different values of $\Delta$: (a) $\Delta=0$,  (b) $\Delta=0.01$, (c) $\Delta=0.04$, (d) $\Delta=0.09$, (e) $\Delta=0.14$, (f) $\Delta=0.18$, (g) $\Delta=0.4$, (h) $\Delta=0.6$.   }
\label{Fig:En0n1}
\end{figure*}
\begin{figure}[tbp ]
\begin{center}
\includegraphics[width=0.5\textwidth]{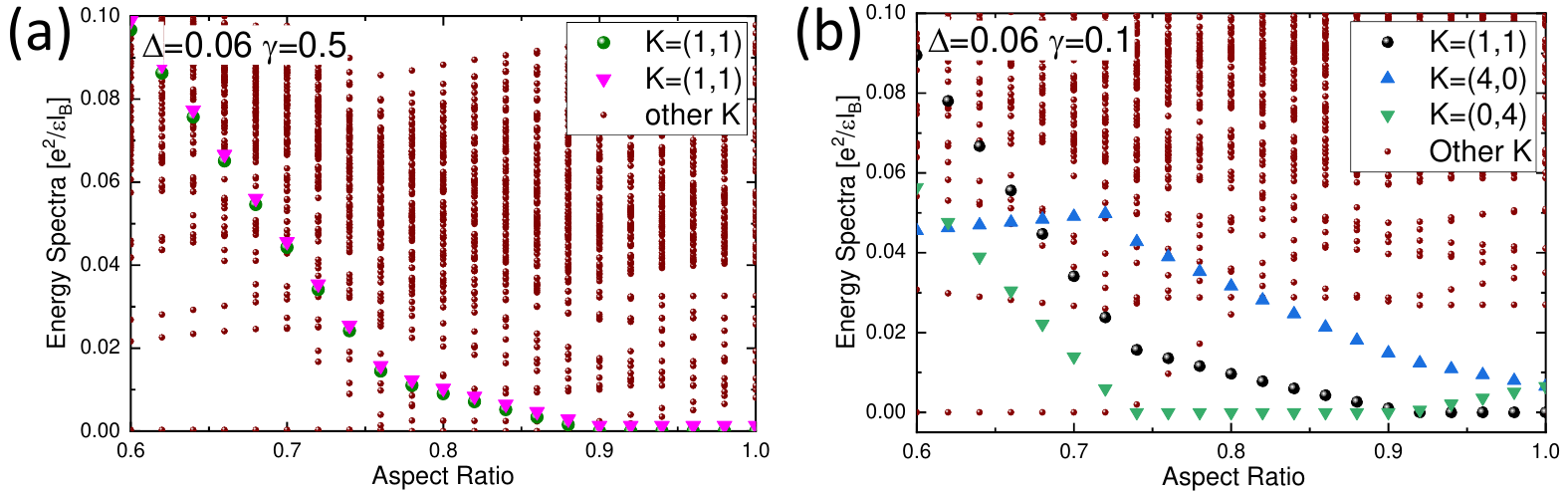}
\end{center}
\par
\renewcommand{\figurename}{Fig.}
\caption{(Color online) The energy spectra as a function of aspect ratio for (a) anisotropic gapless phase (AGP) at $\gamma=0.5$ and (b) CFL state at $\gamma=0.1$. Here, we choose
 $\Delta=0.06$.  }
\label{Fig:Aspect}
\end{figure}

\begin{figure*}[tbp ]
\begin{center}
\includegraphics[width=0.95\textwidth]{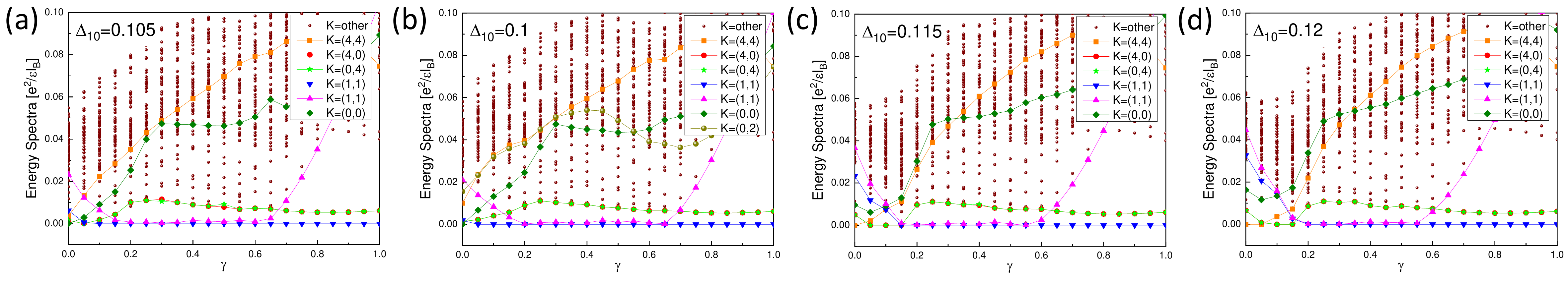}
\end{center}
\par
\renewcommand{\figurename}{Fig.}
\caption{(Color online) The energy spectrum as a function of $\gamma$ for different values of $\Delta$: (a) $\Delta=0.105$,  (b) $\Delta=0.11$, (c) $\Delta=0.115$, (d) $\Delta=0.12$.   }
\label{Fig:Spectra2}
\end{figure*}
\begin{figure*}[tbp ]
\begin{center}
\includegraphics[width=0.8\textwidth]{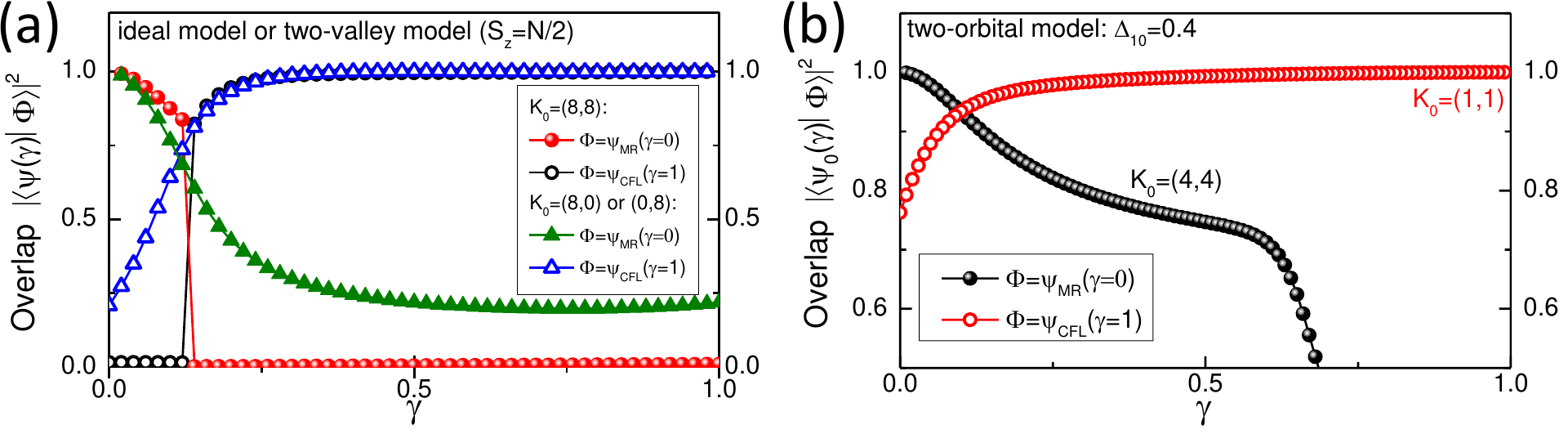}
\end{center}
\par
\renewcommand{\figurename}{Fig.}
\caption{(Color online) The wave function overlap between the MR/CFL state and the state with different $\gamma$. Panel (a) shows the two-valley model at $S_z=N/2$ or the ideal model; Pannel (b) corresponds to the two-orbital model.}
\label{Fig:overlap}
\end{figure*}

\section{Screening effect}
For the BLG, we need to consider the dramatic screening effect on the bare Coulomb interactions. Based on the standard random phase approximation (RPA) approach~\cite{RPA0s,RPA1s}, we take the screening effect into account by replacing the bare Coulomb interaction $V(q)$ with
\begin{equation}
\frac{ V_{\text{SC}}(q) }  { 2\pi e^2/ {\varepsilon l_B}}\approx \frac{{V\left( q \right)}}{{1 + V\left( q \right)\Pi(q)}},
  \end{equation}
 where $V\left( q \right) = \frac{1}{q}\tanh \left( {qd} \right)$ is the Fourier transform of the gate-screened potential with $d$ denoting the distance to graphite gates. The polarization function
 \begin{equation}
 \Pi(q) = a 4 \log(4)\tanh(b q^2 )
    \end{equation}
  is got by a phenomenological method following the approach of Ref.~\cite{RPA2s}. Here $a$ and $b$ are parameters describing phenomenologically the screening in bilayer graphene. If we choose $a$ and $b$ to match the two-band RPA calculation ~\cite{RPA1s,Snizhko2012s}then $b=0.62$ and $a=a_0E_c/\hbar \omega_c$, then we try several values of $a_0$ to study the screening effect. Figure~\ref{Fig:Screen} shows the energy spectra as a function of  $\gamma$ in the sector $S_z=0$ for different $a_0$, it shows that the transition between the MR state and fully polarized CFL state are almost independent on the screening effect. This suggests that the short distance components of the interactions determine  the energetics since the long range part of Coulomb is primarily affected by screening.

\section{Valley symmetry breaking terms.}
 It is by now understood that  short distance corrections to the Coulomb interaction play an important role in graphene~\cite{Kharitonov2012s,Kharitonov2012bs,MacDonald2014s,Fengchengs}. In bilayer graphene  such correction is very natural and simply arises from the finite interlayer distance, which makes the Coulomb repulsion between electrons in the same valley (layer) slightly stronger than that between electrons in opposite valleys (layer)~\footnote{This interaction is responsible for the finite intrinsic interlayer capacitance in BLG, and once accounted one should not add a an extra capacitive cost to the Hamiltonian as this will be double counting such effect.}. The second interaction is an inter-valley swap scattering process. The interactions between particles $i$ and $j$ can be represented by the following valley dependent delta function potential~\cite{Kharitonov2012s,Kharitonov2012bs}:
\be\label{ValleyV}
V_{ij}=\sum_{\mu} V_\mu \tau^i_\mu \tau^j_\mu \delta^{(2)}({\bf r}_i-{\bf r}_j).
\ee
\noindent here $\mu=\{x,y,z\}$, $\tau_\mu$ are Pauli matrices in valley space, and $V_x=V_y=V_\perp$, preserving the global $U(1)$ valley conservation. The coefficient $V_z \sim \pi e^2 d/(\epsilon l^2)$ can be estimated from the Coulomb energy difference in a bilayer. The coefficient $V_\perp$ is harder to estimate microscopically, but it should also scale as $|V_\perp| \sim e^2 a/(\epsilon l^2)$, where $a$ is a short-distance lattice scale. However, both the single component MR and CFL states are {\it hard-core} states, in the sense introduced in Refs.~\cite{MacDonald2014s,Thesiss}, namely their wave-functions are expected to vanish sufficiently fast as any two particles approach each, and, therefore, to a good approximation, they pay no energy under the symmetry breaking terms in Eq.~\eqref{ValleyV}. This means that these interactions will not significantly alter the energetics that we have so far discussed, and, in particular that the pinning for the SU(2) spontaneously valley polarized MR and CFL states is expected to be rather weak.

\section{Two-obital model  with orbital fluctuations}

In the above, we ignored quantum fluctuations between $n=0$ and $n=1$ Landau levels.  In this section, we consider these two Landau levels with different orbital character and the same valley $\psi_{\alpha K}$, $\alpha =\{0,1\}$. The corresponding wavefuctions for $\psi_{0}$ and $\psi_{1}$ orbits are
\be\label{psi1}
\psi_{0}=
\begin{pmatrix}
\phi_{0}\\
0\\
0\\
0
\end{pmatrix}, \ \
\psi_{1}=
\begin{pmatrix}
\sqrt{1-\gamma} \phi_{1}\\
0 \\
\sqrt{\gamma}\phi_{0}\\
0
\end{pmatrix}.
\ee
Then the projected Coulomb interaction given by
\be
\begin{aligned}
V=\sum_{{\mathbf q  \{\alpha\}}}&\frac{v({\mathbf q})}{2A}F_{n_1n_4}({\bf q})F_{n_2n_3}({\bf -q}):\rho_{n_1n_4}^\dagger({\bf q})\rho_{n_2n_3}({\bf q}):\\
&+\Delta_{\mathrm{10}} N_0,
\end{aligned}
\ee
where $\alpha =n_{1,2,3,4} =0,1$ denote orbits. $\Delta_{\mathrm{10}}$  is a single particle spliting between these two orbital flavors. There is no flavor conservation in this case and thus the form factors contain flavor off-diagonal components, and are given by
 \be
\begin{aligned}
 F_{00}({\mathbf q}) &=F_{0}({\mathbf q}), \\
 F_{11}({\mathbf q}) &=(1-\gamma) F_1({\mathbf q})+\gamma F_0({\mathbf q}),\\
 F_{10}({\mathbf q})&=[F_{01}(-{\mathbf q})]^*\\&= \sqrt{1-\gamma} \exp(-{\bf q}^2/4)[-i(q_x+iq_y)/\sqrt{2}].
 \end{aligned}
\ee

We have presented the key results of two-orbital model in the main text, here we present more results  with different parameters to illustrate  the phase diagram in Fig.1(a) in the main text.
Figures~\ref{Fig:En0n1} (a)-(h) show the energy spectra as a function of $\gamma$ with different values of $\Delta_{10}$.  We can see a modest single particle splitting $\Delta_{10}\gtrsim0.2$ is enough to reach the behavior of the ideal limit [see Figs~\ref{Fig:En0n1} (f)-(h)], i. e., the transition between MR state and CFL state takes place at $\gamma\approx 0.15$. The nature of the MR state can be identified by the three-fold degeneracy on top of two-fold center of mass degeneracy in momentum sectors $(K_x,K_y)/(2\pi/N)=(N/2,N/2),(0,N/2),(N/2,0)$. The CFL state in the $\psi_1$ orbit has the ground-state momentum  $\mathbf{K_0}=(1,1)$. At smaller $\Delta_{10}$, we have  CFL-I state, which is the ordinary CFL state realized at half-filled $\psi_0$ orbit, as shown in Figs~\ref{Fig:En0n1} (a)-(d), the ground-state momentum is also $\mathbf{K_0}$.   The reason why CFL-I becomes the ground state near $\Delta_{10}=0$ is that there is an exchange energy gain to occupy $n=0$ orbits due to the smaller spatial extension and hence larger exchange holes, and therefore at half-filling the state is the conventional $n=0$ CFL. As $\gamma\rightarrow1$ the $\psi_1$ Landau level becomes effectively an $n=0$ level and thus we have an SU(2) invariant two-component unpolarized CFL near $\gamma\rightarrow1$ and $\Delta_{10}\rightarrow0$  [see Figs~\ref{Fig:En0n1} (a)-(b)], which has ground-state momentum $\mathbf{K_0}=(4,4)$. We also note that in the region near the $n=1/0$ level crossing around $\Delta_{10}\sim 0.11 e^2/\epsilon l_B$ and $\gamma \lesssim 0.15$ where many phases meet, as shown in Fig.~\ref{Fig:Spectra2}. It is possible that other phases might be present in this narrow parameter regime, such as $221$ parton state~\cite{Wu2017s}, which we hope future studies will further address this interesting possibility.

In particular, for the intermediate $\Delta_{10}$, we find a new gapless phase, as shown in  Figs~\ref{Fig:En0n1} (c)-(e). This phase is characterized by a multiplicity of low-lying states and a robust ground state quasi-degeneracy, indicating a gapless symmetry breaking state that we label anisotropic gapless phase (AGP) in the phase diagram. The energy spectra of such gapless phase [see Fig.~\ref{Fig:Aspect}(a)] has large sensitivity to aspect ratio variation, in contrast to the nearby CFL phase [see Fig.~\ref{Fig:Aspect}(b)]. Moreover, we find no evidence of quasi-degenerate feature in energy spectra among different momentum sectors, thus it is unlikely to be a stripe phase, although it could be that our system sizes are too small to rule it out.  Since we have evidence of broken rotational symmetry, i.e., the quasi-degenerate spectra and high sensitivity to aspect ratio [see Fig.~\ref{Fig:Aspect}],we have the possibility of a nematic state. We have also verified that the charge gap of this state is not substantially different from the neighboring composite fermi liquid states indicating that it could be a compressible phase. We hope to further clarify its nature in future studies.

 In the limit of large $|\Delta_{10}|$, the two-orbital model reduces to that of a single half-filled Landau level, and hence it has particle-hole symmetry for any two-body interaction. Therefore, in this limit, there is no energetic distinction between MR Pfaffian and anti-Pfaffian.  However, at finite $\Delta_{10}$, there is a mixing of $n=1/0$ orbits that would split such degeneracy of the MR Pfaffian and anti-Pfaffian in the thermodynamic limit. For all the systems we studied we observe a rather smooth evolution of the spectrum up to  $\Delta_{10}\sim 0.2 e^2/\epsilon l_B$. This indicates that there is no change from MR Pfaffian to anti-Pfaffian as function of $\Delta_{10}$ up to $\Delta_{10}\sim 0.2 e^2/\epsilon l_B$. Therefore, there is presumably a uniquely chosen state as function of $\Delta_{10}$, at least up to $\Delta_{10}\sim 0.2 e^2/\epsilon l_B$. The state is however expected to change for positive vs negative $\Delta_{10}$, namely, if the MR Pfaffian state is favored for positive  $\Delta_{10}$ the anti-Pfaffian would be favored at negative  $\Delta_{10}$. This follows from the fact that a global particle-hole conjugation involving both the n=0 and n=1 LL would change the sign of  $\Delta_{10}$ and map MR Pfaffian to anti-Pfaffian.  Previous numerical studies have found that the MR Pfaffian state tends to be favored when only the n=0 and n=1 LL are kept and one has the usual energetic ordering of GaAs, namely for positive $\Delta_{10}$~\cite{Rezayi2017s}. Therefore, from these results, we can speculate that the MR Pfaffian would be favored for the case of our ideal Hamiltonian when $\Delta_{10}>0$ and the anti-Pfaffian tends to be favored when $\Delta_{10}<0$. We caution, however that this competition is so delicate, that in GaAs there is a reversal of which of the two states is favored when more than two Landau levels are kept, leading to the expectation that the anti-Pfaffian is favored~\cite{Rezayi2017s}. Therefore, it would be very hard at the moment for us to make any reliable predictions for which of the two is favored in realistic BLG, since the LL mixing from higher Landau levels would be different from the GaAs case. We notice that although BLG tends to have a positive $\Delta_{10}$, it might possible to change the sign of $\Delta_{10}$ by inducing a sufficiently large interlayer bias~\cite{Inti2019s}.

\begin{figure*}[tbp]
\begin{center}
\includegraphics[width=0.9\textwidth]{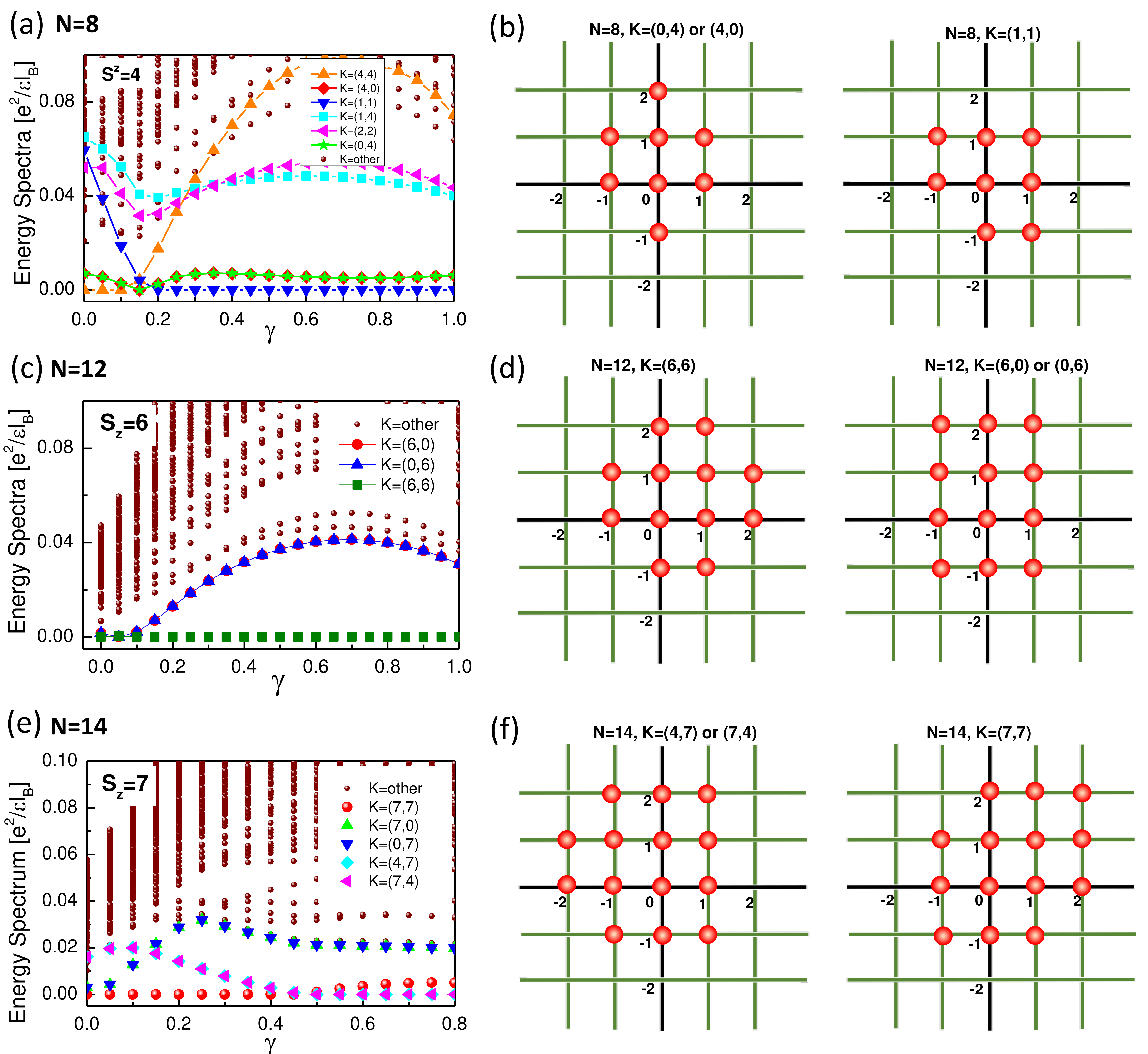}
\end{center}
\par
\renewcommand{\figurename}{Fig.}
\caption{(Color online) Panels (a), (c), (e) show the energy spectra as a function of  $\gamma$ for the two-valley model at $S_z=N/2$ or the ideal model for systems with $N=8$ (a), $N=12$ (c), $N=14$ (e).
Panels (b), (d), (f) show the  structure of the CFL clusters that are found to have the lower mean field energy for the two-valley model at $S_z=N/2$ or the ideal model for systems with $N=8$ (b), $N=12$ (d), $N=14$ (f). Here we should note panel (e) is the same as Fig.~\ref{Fig:N14N16}(a), we plot here again for comparison and completeness.}
\label{Fig:TransitionType}
\end{figure*}

\begin{figure*}[tbp]
\begin{center}
\includegraphics[width=0.9\textwidth]{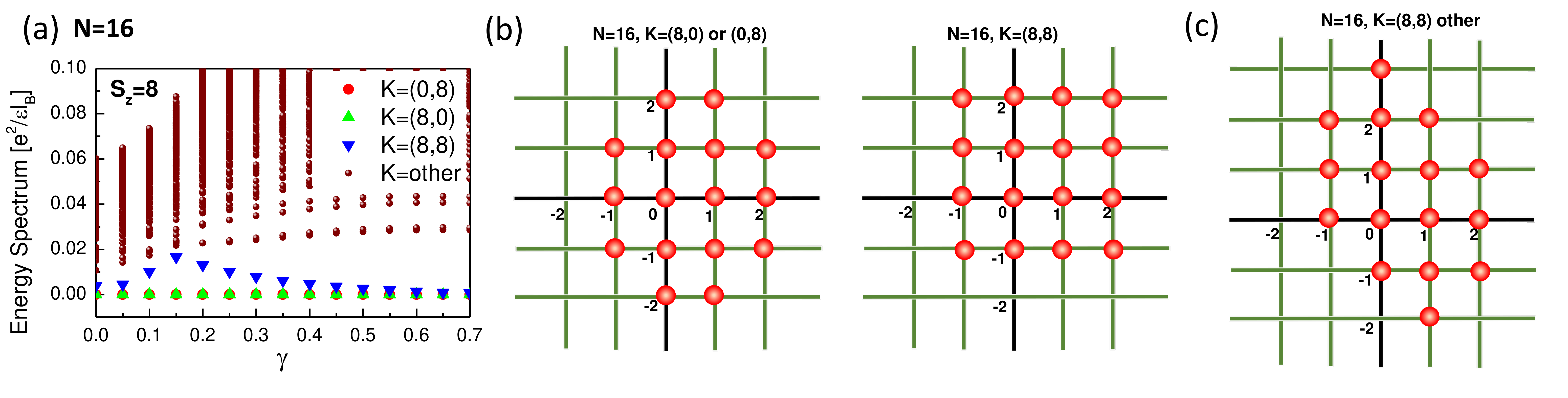}
\end{center}
\par
\renewcommand{\figurename}{Fig.}
\caption{(Color online) Panels (a) shows the energy spectra as a function of  $\gamma$ for the two-valley model at $S_z=N/2$ or the ideal model for systems with $N=16$.
Panels (b) shows the corresponding  structure of the CFL clusters that are found to have the lower mean field energy. Panels (c) shows the structure of the CFL clusters in K=(8,8) sector but with a mean field energy difference of $10\%$ with respect to the cluster in (b).  Here we should note panel (a) is the same as Fig.~\ref{Fig:N14N16}(b), we plot here again for comparison and completeness.}
\label{Fig:TransitionType2}
\end{figure*}

\begin{figure*}[tbp]
\begin{center}
\includegraphics[width=0.9\textwidth]{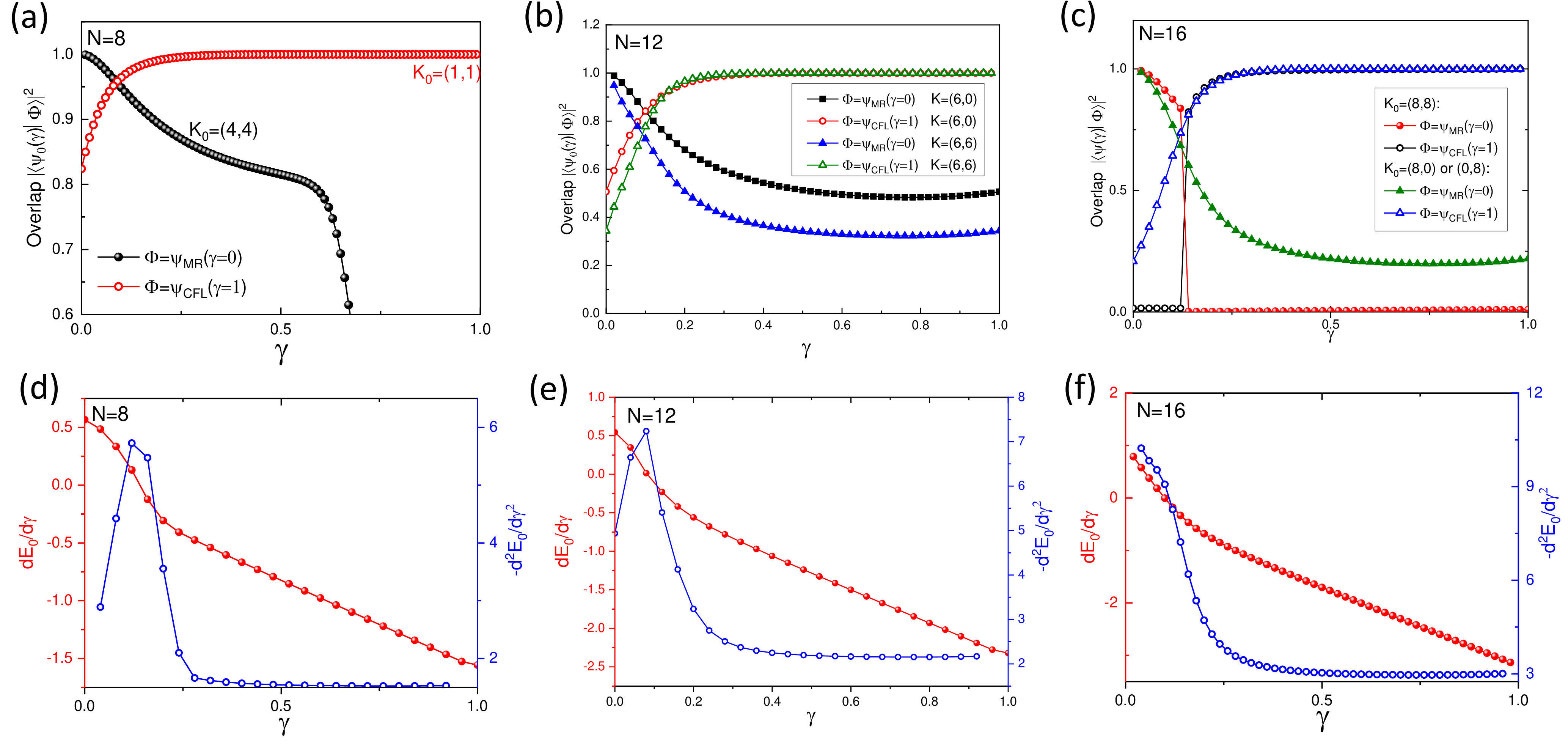}
\end{center}
\par
\renewcommand{\figurename}{Fig.}
\caption{(Color online) {Panel (a-c) show the wave function overlap between the MR/CFL state and the state with different $\gamma$ for the two-valley model at $S_z=N/2$ or the ideal model with $N=8$ (a), $N=12$ (b), $N=16$ (c). Panel (d-f)  show the first (red curve) and second order (blue curve) derivatives of the ground-state energy for the two-valley model at $S_z=N/2$ or the ideal model for systems with $N=8$ (d), $N=12$ (e), $N=16$ (f). Here we should note panel (c) is the same as Fig.~\ref{Fig:overlap}(a), we plot here again for comparison and completeness.}}
\label{Fig:TransitionType3}
\end{figure*}

\section{Wave Function Overlap}
In the main text, the MR state is mainly identified from the degeneracy and its corresponding degenerate sectors, the CFL state is identified by the quantum number of the ground state after correctly capturing the corrections on finite sized torus~\cite{HR2000s,Zheng2018s,RR1994s,Read1994s}, both of which have been proved to successfully capture these two kinds of state on torus with different electron numbers. In this section, we performed the wave-function overlap calculations to further support the conclusions in the main text. The wave function overlap is calculated between the MR/CFL state with the ground state at different $\gamma$, as shown in Fig.~\ref{Fig:overlap}, we find that, in both the ideal model, the two-valley model and the two-orbital model, the squared overlap indicate the state at smaller $\gamma$ is indeed MR state, while the state at larger $\gamma$ is CFL state. In addition, the intersection of different curves is consistent with the phase transitions identified from the level crossing in energy spectra.

\section{The type of  MR to CFL transition}
Our exact diagonalization study is limited to systems up to 16 particles. This combined with the fact the the CFL is a gapless state with a Fermi surface, which leads to enhanced finite size effects, makes it challenging to unambiguously determine whether the transition is continuous or discontinuous in the thermodynamic limit even for our ideal model involving a single Landau level. However, there are some salient features that suggest the transition in the ideal limit is continuous. As we have discussed, for an even number of electrons the Moore-Read state always displays three topological copies of the ground state at specific many-body momenta
$(K_x,K_y)=(N/2,N/2),(0,N/2),(N/2,0)$. The momenta of the low lying states of the CFL however change with
system size due the shell filling effect, according to the rules discussed in Refs.~[\onlinecite{Haldane1985s,HR2000s,Zheng2018s,Read1994s}] and summarized in Section~\ref{Sec:Two-valley} of the supplementary material. We have noticed however that in almost all cases whenever there is a finite size CFL Fermi sea cluster that has low energy and the composite fermion momenta are occupied in pairs related by a $k\rightarrow-k$ symmetry, such state will smoothly evolve into one of the ground states of the Moore-Read state as a function of the
tuning parameter $\gamma$. We will illustrate below this behavior for system sizes N={8,12,14}.

For N=8 system,  the spectrum is shown in Fig.~\ref{Fig:TransitionType} (a). We have identified the lowest mean field energy CFL clusters to agree with the above ED results and have the structure illustrated in  Fig.~~\ref{Fig:TransitionType} (b). Here the red balls are the occupied momenta by the composite fermions. Notice that the the K=(0,4) and (4,0) cluster has $k\rightarrow-k$ symmetry, whereas the K=(1,1) cluster does not. We see that the K=(0,4) and (4,0) CFL clusters become Moore-Read ground states whereas the $K=(1,1)$ cluster becomes an excited state in the MR phase. This is consistent with the idea that the clusters with $k\rightarrow-k$ symmetry evolve smoothly into paired states.

For N=12 system, the spectrum is shown in Fig.~~\ref{Fig:TransitionType}(c). The CFL clusters that are found to have the lowest mean field energy have the structure depicted in  Fig.~~\ref{Fig:TransitionType}(d). We see that in this case all the lowest energy CFL clusters have $k\rightarrow-k$ symmetry and have momenta that correspond to the ground states of MR, and we see that they evolve smoothly into the MR ground states.

For N=14 system, the spectrum is shown in Fig.~~\ref{Fig:TransitionType}(e). The CFL clusters that are found to have the lowest mean field energy have the structure depicted in  Fig.~~\ref{Fig:TransitionType}(f).  We see that the CFL clusters with momenta K=(4,7) and (7,4) do not have $k\rightarrow-k$ symmetry and they do not evolve into the topological ground state sectors of the MR, whereas the cluster K=(7,7) has $k\rightarrow-k$ symmetry and evolves smoothly into a corresponding ground state of the MR state.

We have only encountered a deviation from this pattern for N=16 particles. Below we illustrate the spectrum [see Fig.~\ref{Fig:TransitionType2}(a)] and the lowest energy CFL clusters [see Fig.~\ref{Fig:TransitionType2}(b)] for this case. We see that in this case all the lowest energy CFL clusters have  $k\rightarrow-k$ symmetry and have momenta that correspond to the ground states of MR, as shown in Fig.~\ref{Fig:TransitionType2}(b). However, in the spectrum we see that the K=(8,8) CFL state does not evolve smoothly into the corresponding K=(8,8) state in the MR side. We do not understand the detailed origin of this deviation, but presume that it is related to the fact that for N=16 there are many more low lying states with the same momentum as the corresponding CFL. For example, we have encountered another cluster shown in Fig.~\ref{Fig:TransitionType2}(c) with the same momentum and a mean field energy difference of $10\%$ with respect to the cluster shown in Fig.~\ref{Fig:TransitionType2}(b), indicating that mixing between different clusters becomes more important as the system size grows.

We have also computed overlaps and derivatives of the ground-state energy as a function of $\gamma$  described above and they all generically present display a smooth behavior near
the transition between the CFL and the MR state, as shown in Figs.~\ref{Fig:TransitionType3}, indicating the transition is continuous. The only case where we have encountered an abrupt drop of the overlap is again for the K=(8,8) sector for N=16 particles that we described above, and it is shown in Fig.~\ref{Fig:TransitionType3}(c). We suspect this is an isolated anomaly associated with the competing clusters in this case near the transition leading to a weakly avoided level crossing, but overall the evidence from all other cases indicates a rather smooth transition from CFL into MR.  We should also note that the derivatives of the ground-state energy is computed by  the lowest energy for each $\gamma$, for $N=16$ system, the $K= (8,0)$ and $(0,8)$ sectors have the lowest energy, which also have the continuous variations of the overlap as shown in Fig.~\ref{Fig:TransitionType3}(c). In this work we have shown the BLG would a suitable platform to realize a clean phase transition between MR and CFL, we hope that experiments in BLG will ultimately shed light on the nature of the transition in the thermodynamic limit.


\begin{thebibliography}{99}
\bibitem{footnote1}We only show states in the upper quarter of the many-body Brillouin zone since the remainder are related by point group symmetry.
\end{thebibliography}
\end{document}